%% file: main.tex
\documentclass[journal,12pt,onecolumn,draftclsnofoot,comsoc]{IEEEtran}
\input{header}

\title{MU-MIMO Uplink Timely Throughput Maximization for Extended Reality Applications}
\input{authorinfo}

\begin{document}

\maketitle
\input{abstract}
\textbf{Preprint Notice:} This work has been submitted to the IEEE for possible publication. Copyright may be transferred without notice, after which this version may no longer be accessible.

\input{introduction}
\input{networkmodel}
\input{problemformulation}
\input{proposedsolution}
\input{numericalresults}

\input{conclusion}
\bibliographystyle{unsrt}

\bibliography{references}
\end{document}

%% file: header.tex
\usepackage{ifpdf}
\usepackage{comment}
\usepackage{textcomp}
\usepackage{stfloats}
\usepackage{adjustbox}

\usepackage{epsfig}
\usepackage{graphics}
\usepackage{subcaption}
\usepackage[caption=false,font=normalsize,labelfont=sf,textfont=sf]{subfig}

\usepackage{amsmath}
\usepackage{amsfonts}
\usepackage{amssymb}  
\usepackage{amsthm}
\usepackage{mathtools}
\usepackage{bbm}
\usepackage{empheq}
\usepackage{cases}
\usepackage[acronym]{glossaries}
\usepackage{algorithm}
\usepackage{algorithmicx}
\usepackage{algpseudocode}

\algnewcommand\algorithmicinput{\textbf{Input:}}
\algnewcommand\INPUT{\item[\algorithmicinput]}
\algrenewcommand\algorithmicrequire{\textbf{Initialize:}}
\algnewcommand\INIT{\item[\algorithmicrequire]}
\algblockdefx{MRepeat}{EndRepeat}{\textbf{repeat}}{}
\algnotext{EndRepeat}
\algrenewcommand\algorithmicforall{\textbf{for each}}

\usepackage{array}

\usepackage{cite}

\usepackage{url}

\usepackage{xcolor}
\usepackage{enumitem}

\usepackage{fancyhdr}

\theoremstyle{definition}

\DeclareMathAlphabet\mathcallite{OMS}{cmsy}{m}{n}
\DeclareMathAlphabet\mathbfcal{OMS}{cmsy}{b}{n}

\DeclarePairedDelimiter\ceil{\lceil}{\rceil}

\DeclareMathOperator*{\maximize}{maximize}

\newacronym{3GPP}{3GPP}{third generation partnership project}
\newacronym{UE}{UE}{user equipment}
\newacronym{DU}{DU}{distributed unit}
\newacronym{DUs}{DUs}{distributed units}
\newacronym{CU}{CU}{centralized unit}
\newacronym{CSI}{CSI}{channel state information}
\newacronym{MIMO}{MIMO}{multiple input multiple output}
\newacronym{MU-MIMO}{MU-MIMO}{multi-user multiple input multiple output}
\newacronym{SU-MIMO}{SU-MIMO}{single-user MIMO}
\newacronym{QoS}{QoS}{quality of service}
\newacronym{SIC}{SIC}{successive interference cancellation}
\newacronym{MDP}{MDP}{Markov Decision Process}
\newacronym{CMDP}{CMDP}{constrained Markov Decision Process}
\newacronym{SNR}{SNR}{signal-to-noise ratio}
\newacronym{DL}{DL}{downlink}
\newacronym{UL}{UL}{uplink}
\newacronym{TDMA}{TDMA}{Time Division Multiple Access}
\newacronym{5G}{5G}{fifth generation}
\newacronym{5G-A}{5G-A}{5G-Advanced}
\newacronym{6G}{6G}{sixth generation}
\newacronym{SE}{SE}{spectral efficiency}
\newacronym{EE}{EE}{energy efficiency}
\newacronym{FDMA}{FDMA}{Frequency Division Multiple Access}
\newacronym{TDD}{TDD}{time division duplex}
\newacronym{FDD}{FDD}{frequency division duplex}
\newacronym{NUM}{NUM}{network utility maximization}
\newacronym{MCS}{MCS}{modulation and coding scheme}
\newacronym{PDB}{PDB}{packet delay budget}
\newacronym{AoI}{AoI}{age of information}
\newacronym{RL}{RL}{reinforcement learning}
\newacronym{DNN}{DNN}{deep neural network}
\newacronym{DRL}{DRL}{deep reinforcement learning}
\newacronym{XR}{XR}{extended reality}
\newacronym{AR}{AR}{augmented reality}
\newacronym{MR}{MR}{mixed (augmented) reality}
\newacronym{VR}{VR}{virtual reality}
\newacronym{PRB}{PRB}{physical resource block}
\newacronym{RB}{RB}{resource block}
\newacronym{ADU}{ADU}{application data unit}
\newacronym{TB}{TB}{transport block}
\newacronym{BSR}{BSR}{buffer status report}
\newacronym{RE}{RE}{resource element}
\newacronym{TTI}{TTI}{transmission time interval}
\newacronym{OFDMA}{OFDMA}{orthogonal frequency division multiple access}
\newacronym{OFDM}{OFDM}{orthogonal frequency division multiplexing}
\newacronym{SINR}{SINR}{signal-to-interference-plus-noise ratio}
\newacronym{QoE}{QoE}{quality of experience}
\newacronym{AoT}{AoT}{age of transmission}
\newacronym{RAN}{RAN}{radio access network}
\newacronym{CMD}{CMD}{correlation matrix distance}
\newacronym{PAoT}{PAoT}{peak age of transmission}
\newacronym{PAoI}{PAoI}{peak AoI}
\newacronym{LM}{LM}{Lagrange Multiplier}
\newacronym{MAC}{MAC}{medium access control}
\newacronym{PHY}{PHY}{physical}
\newacronym{LA}{LA}{link adaptation}
\newacronym{TPM}{TPM}{transition probability mapping}
\newacronym{MU}{MU}{multi-user}
\newacronym{AWGN}{AWGN}{additive white Gaussian noise}
\newacronym{2D}{2D}{two-dimensional}
\newacronym{3D}{3D}{three-dimentsional}
\newacronym{LSTM}{LSTM}{long short-term memory}
\newacronym{gNB}{gNB}{gNodeB}
\newacronym{RNN}{RNN}{recurrent neural network}
\newacronym{BiRNN}{BiRNN}{bi-directional recurrent neural network}
\newacronym{ICI}{ICI}{inter-cell interference}
\newacronym{NGWN}{NGWN}{next generation wireless networks}
\newacronym{HMD}{HMD}{head-mounted display}
\newacronym{MU-MR}{MU-MR}{multi-user MR}
\newacronym{AA}{AA}{application awareness}
\newacronym{Qo(S)E}{Qo(S)E}{QoS/QoE}
\newacronym{PF}{PF}{proportional fair}
\newacronym{eMBB}{eMBB}{enhanced mobile broadband}
\newacronym{PDU}{PDU}{protocol data unit}
\newacronym{IP}{IP}{Internet protocol}
\newacronym{KPI}{KPI}{key performance indicator}
\newacronym{LMMSE}{LMMSE}{linear minimum mean sqaure error}
\newacronym{TD}{TD}{time-domain}
\newacronym{TRX}{TRX}{transceiver chains}
\newacronym{AM}{AM}{arithmetic mean}
\newacronym{GM}{GM}{geometric mean}
\newacronym{pmf}{pmf}{probability mass function}

%% file: authorinfo.tex
\author{\IEEEauthorblockN{Ravi Sharan B A G, K. Pavan Srinath, Alvaro Valcarce Rial, and Baltasar-Beferull Lozano}
\thanks{
    Ravi Sharan is with Nokia Bell Labs, Germany. K. P. Srinath and Alvaro V. Rial are with Nokia Bell Labs, France. Email: \{ravi.bhagavathula, pavan.koteshwar\_srinath, alvaro.valcarce\_rial\}@nokia-bell-labs.com. Baltasar B. Lozano is with SimulaMet and University of Agder, Norway. Email : baltasar@simula.no. Part of this work was carried out when Ravi Sharan was enrolled as a PhD student at the University of Agder, Norway.
}
}

%% file: abstract.tex
\begin{abstract}
    In this work, we study the cross-layer timely throughput maximization for extended reality (XR) applications through uplink multi-user \acrshort{MIMO} (MU-MIMO) scheduling. Timely scheduling opportunities are characterized by the peak age of information (PAoI)-metric and are incorporated into a network-side optimization problem as constraints modeling user satisfaction. The problem being NP-hard, we resort to a signaling-free, weighted proportional fair-based iterative heuristic algorithm, where the weights are derived with respect to the PAoI metric. Extensive numerical simulation results demonstrate that the proposed algorithm consistently outperforms existing baselines in terms of \gls{XR} capacity without sacrificing the overall system throughput.
\end{abstract}


%% file: introduction.tex
\section{Introduction}

\Gls{XR} services encompassing \gls{MR} and \gls{VR} technologies are shaping the future immersive multimedia applications. A seamless \gls{XR} experience over wireless cellular networks involve delivering high and reliable data rates ($\approx 30-150$ Mb/s) with low latencies ($\approx 5-30$ ms). These requirements are further amplified in co-located multi-user \gls{MR} scenarios, where users simultaneously interact with one another and their virtual environments. Regardless, \gls{MR} applications are equally intensive in the \gls{UL} as they are in the \gls{DL}. This is because a \gls{UE} continuously transmits raw video for 3D-rendition at the \gls{RAN}~\cite{vpetrov2022xrover5g}. Thus, \gls{UL} optimization plays a crucial role in realizing the full potential of these \gls{XR} services.

For delivering a guaranteed \gls{QoS} in the \gls{UL}, \gls{UE} scheduling with delay-specific information is currently being considered in the \gls{3GPP} studies~\cite{3gpprel19sa2}. \Gls{QoE} is handled separately by extracting application awareness at the video frame-level and embedding it into the \gls{MAC} layer as packet-level metadata~\cite{aamiri2024applicationawareness}. On the other hand, with latest advances in the \gls{PHY} layer, \gls{MU-MIMO} is being seriously investigated for \gls{6G} networks~\cite{jandrews20246gtakeshape}. \gls{UE} scheduling in \gls{MU-MIMO} becomes extremely challenging due to intra-cell interference, which impacts precoding at the \gls{PHY} layer and consequently the \acrshort{Qo(S)E}. Thus, a careful consideration of both \gls{PHY}-\gls{MAC} aspects is imperative for \gls{MU-MIMO} \gls{UL} scheduling in \gls{XR} applications.

Scheduling schemes based on the well known \gls{PF} metric are widely adopted in the commercial \gls{5G} cellular networks~\cite{yhao2021muxurllc}. The \gls{PF} metric aims to maximize the system throughput based on the channel conditions while maintaining long-term fairness among the \glspl{UE}. Going forward, \acrshort{Qo(S)E}-related extensions to the \gls{PF} metric are necessary to successfully employ it in \gls{XR} scenarios. However, acquiring the aforementioned application awareness information involves additional signaling overhead in the \gls{UL} and cannot be fully generalized across applications. \textit{In this work, we follow a signaling-free approach and employ \gls{AoI} alongside \gls{PF} metric for \gls{MU-MIMO} \gls{UL} \gls{UE} scheduling in a \gls{XR} scenario}. The \gls{AoI} is defined as the time elapsed since a \gls{UE}'s last generated successful packet, where it jointly captures the inter-delivery times and the packet-delays~\cite{rdyates2021aoisurvey}. Furthermore, we use the \gls{PAoI} metric to model the \acrshort{Qo(S)E} of the \glspl{UE}, since it can be easily coupled with the \gls{PDB}~\cite{skumar2021onlineenergy}. To the best of our knowledge, we are not aware of existing works on \gls{AoI}-aware \gls{UL} scheduling with \gls{XR} traffic considering full-blown \gls{MU-MIMO} \gls{PHY} aspects.

\textbf{Related Literature:} Current literature on \gls{XR} \gls{UE} scheduling heavily focuses on \gls{DL} \gls{SU-MIMO} scenarios. For instance, the work in~\cite{ppouria2024pduset} proposed a \gls{QoS}-aware scheduling heuristic algorithm for \gls{XR} and \gls{5G} \acrshort{eMBB} traffic. Here, the scheduling metric is modeled with respect to the recently introduced \gls{PDU}-set concept. The authors in~\cite{cerkai2021framelevel} propose two scheduling heuristic algorithms to maximize the number of satisfied \gls{XR} \glspl{UE} by combining multiple video frames into one entity. The work in~\cite{bbiljana2023xrloopback} proposed a signaling-aided scheme to meet the target \acrshort{Qo(S)E} and \gls{XR} \glspl{KPI}, where the \gls{gNB} collects periodic measurements from the \glspl{UE}. Above schemes exploit packet-level metadata to model the \gls{DL} scheduling metrics, which can be hard to incorporate in \gls{UL} \gls{UE} scheduling without introducing additional signaling overhead. Firstly, the signaling overhead can have a detrimental effect on the achievable \acrshort{Qo(S)E} under stringent latency requirements of the \gls{XR} applications. Even if the signaling is well-handled, intra-cell interference poses an additional challenge for \gls{UE} scheduling in \gls{MU-MIMO}. 

On the other hand, \gls{AoI}-aware \gls{UL} scheduling for short-packet communications with single-antenna \glspl{UE} are explored in~\cite{ccwu2021aoiaware5g} and~\cite{sli2022aotoptimalscheduling}. The work in~\cite{ccwu2021aoiaware5g} proposes a \gls{DRL} solution for throughput maximization subject to network-wide sum-\gls{AoI} constraint in a \gls{SU-MIMO} setting. Finally, an interference-approximated whittle-index policy is proposed to minimize the network-wide sum-\gls{AoI} subject to instantaneous deadline in~\cite{sli2022aotoptimalscheduling}. However, these works do not shed light on the \gls{AoI} behavior under \gls{XR} traffic. Within this context, we now present the system model, which is followed by problem formulation, proposed solution and numerical results.

%% file: networkmodel.tex
\section{System Model}
\label{sec.networkmodel}

We consider a \gls{MU-MIMO} scenario with a single \gls{gNB} serving $N$ \gls{XR} \glspl{UE} in the \gls{UL}. Let $\mathcal{N}:=\{1,2,\ldots,\Bar{N}\},~|\mathcal{N}| = \Bar{N}$ denote the set of all \glspl{UE} served by the \gls{gNB}. A subset of \glspl{UE} from $\mathcal{N}$ are multiplexed together during a single transmission opportunity in the same time-frequency \acrshort{2D} resource grid. In the \gls{TD}, the transmission opportunity is represented by \gls{TTI} $t \in \mathcal{T} \subseteq \mathbb{N}^{+}$. A \gls{TTI} comprises $N_{sym}$ \gls{OFDM} symbols. In the frequency axis, there are $F$ \glspl{RB} with each \gls{RB} $f \in \mathcal{F} := \{1,2,...,F\}$ spanning $N_{sc}$ sub-carriers. Thus, for every \gls{TTI} $t$, there are a total of $I_f = N_{sc} \cdot N_{sym}$ \glspl{RE} per each \gls{RB} $f$. Here, a \gls{RE} is the smallest unit of the \acrshort{2D} resource grid. Furthermore, we assume that the \gls{gNB} operates with $N_G$ \gls{TRX} and a \gls{UE} $n, \forall n \in \mathcal{N}$ is equipped with $N_U$ \gls{TRX} w.l.o.g. At any \gls{TTI} $t$, a \gls{UE} $n$ operates with $\lambda_n(t) \leq N_U$ spatial streams of data. Consequently, the total number of spatial streams from the set of co-scheduled \glspl{UE}, $\mathcal{N}^{\prime}(t) \subseteq \mathcal{N}$ result in $\Lambda(t) = \sum_{n \in \mathcal{N}^{\prime}(t)}\lambda_n(t)$. Moreover, there is an upper limit of $\Lambda(t) \leq \Bar{\Lambda} \leq N_G$ on the total number of spatial layers from $\mathcal{N}^{\prime}(t)$ at any \gls{TTI} $t$. 

As for the wireless channel, we assume a flat and block fading channel, i.e., it remains constant for a single \gls{TTI} $t$ and varies across different \glspl{TTI} in a correlated manner. Similarly, the channel is assumed to remain constant over a single \gls{RB} $f$ and vary across different \glspl{RB}. The received complex signal vector $\mathbf{y}(i_f;t) \in \mathbb{C}^{N_G \times 1}$ at the \gls{gNB}, at \gls{TTI} $t$ and on the \gls{RE} $i_f \in \mathcal{I}_f:=\{1,2,...,I_f\}$ is given by:
\begin{align}\label{eq.receivedsignal}
    &\mathbf{y}(i_f;t) \coloneqq\sum_{n \in \mathcal{N}^{\prime}(t)}\mathbf{H}_{n}(f;t)\mathbf{x}_{n}(i_f;t) + \pmb{\eta}(i_f;t).
\end{align}
In the above expression, $\mathbf{H}_{n}(f;t) \in \mathbb{C}^{N_G \times L_{n}(t)}$ is the effective \gls{UL} channel between the \gls{UE} $u_{n}$ and the \gls{gNB}. The effective channel is assumed to be estimated using known pilot symbols and subsumes the precoding operation. The term $\mathbf{x}_{n}(i_f;t) \in \mathbb{C}^{L_{n}(t) \times 1}$ is the data vector of \gls{UE} $u_{n}$ that is transmitted on the \gls{RE} $(i_f;t)$. Finally, $\pmb{\eta}(i_f;t) \in \mathbb{C}^{N_G \times 1}$ collectively denotes the inter-cell interference (if any) plus thermal noise with mean $\pmb{0}$ and covariance matrix $\mathbf{R}_{cov} \in \mathbb{C}^{N_G \times N_G}$. Notice that the index for the effective channel matrix is with respect to $f$, compared to the remaining terms, which is a consequence of the assumptions on the channel above. The post equalization effective \gls{SINR} estimate of a \gls{UE} $n$ on \gls{RE} $i_f$, denoted by $\rho_n(i_f;t)$, is obtained as a function of $\mathbf{H}_{n}(f;t)$, $\mathbf{H}_{n^{\prime}}(f;t), \forall {n^{\prime}} \in \mathcal{N}^{\prime}(t)\setminus\{n\}$ and $\mathbf{R}_{cov}$ (expression omitted here due to space constraints, see~\cite[Eq. (5)]{kpsrinath2024joint}). The \textit{achievable throughput} of a \gls{UE} $n$, denoted by $Q_{n}(t) \in \mathbb{R}_{\geq 0}$, can then be approximated~\cite{kpsrinath2024joint} as follows:
\begin{align}\label{eq.uethroughput}
    &Q_n(t) \leq \lambda_n(t)\cdot \biggl(\sum_{f \in \mathcal{F}} \sum_{i_f \in \mathcal{I}_f} \log_2\bigl(1+\hat{\rho}_n(i_f;t)\bigr)\biggr),
\end{align}
where, $\hat{\rho}_n(i_f;t)$ represents the transmit power scaled \gls{SINR}.

Following sequence of operations ensue at the \gls{gNB}. At the beginning of a \gls{TTI} $t$, all the associated \glspl{UE} send their \gls{BSR} $b_{n}(t) \in \mathcal{B} := [0,\Bar{B}_n]$ to the \gls{gNB} indicating the number of bits pending transmission. Here, $\Bar{B}_n$ denotes the maximum allowable size of a packet. The term packet is used in this work to refer to a data-burst in \gls{3GPP} \gls{XR} specifications, which comprises multiple \glspl{PDU} grouped together with respect to the underlying video frames with periodic arrivals. The \gls{gNB} then selects a subset of \glspl{UE} and for each $n \in \mathcal{N}^{\prime}(t)$ indicates the scheduling decision using a binary variable $\beta_{n}(t)=1$. The \gls{gNB} also assigns the number of spatial streams, $\lambda_{n}(t)$ and the \gls{MCS} value to be used by the co-scheduled \glspl{UE} in $\mathcal{N}^{\prime}(t)$. The co-scheduled \glspl{UE} then perform \gls{RB} selection, \gls{UL} power control and transmit a fraction of the packet by means of \glspl{TB}. The \gls{TB} bit-size at a \gls{UE} is proportional to its assigned \gls{MCS} value. The \gls{gNB} sends back an instantaneous feedback $s_n(t) \in \{0,1\}$, upon successful/unsuccessful reception of the corresponding \glspl{TB}. Note that we do not consider retransmissions in this work. The \glspl{UE} update their \gls{BSR} based on the $s_n(t)$ value and the process repeats over time.

Following packet generation and \gls{AoI} modeling is considered in this work. A new packet of length $B_n,~\text{s.t.}~B_n \leq \Bar{B}_n$ is generated at a \gls{UE} $n$ only when the current packet:
\begin{enumerate}[label={(\alph*)}]
    \item is either successfully received at the \gls{gNB} or,
    \item exceeds the \gls{PDB}, $\Bar{D}$ leading to a packet failure. 
\end{enumerate}
Additionally, we assume that $B_n > Q_n(t), \forall n \in \mathcal{N}, \forall t \in \mathcal{T}$, implying that a \gls{UE} takes several \glspl{TTI} to successfully transmit its packet. A \gls{UE}'s \gls{AoI} value, denoted by $\Delta_{n}(t) \in \mathbb{N}^{+}$, increases linearly with time and is reset to its default value (set to one in this work w.l.o.g) only when the \gls{UE} is scheduled for transmission and the current packet is successfully transmitted. We also consider a finite-value clipping of the \gls{AoI} beyond $\Bar{\Delta}$, which is chosen to be greater than $\Bar{N}$ and $\Bar{D}$. The \gls{gNB} maintains a separate counter for each \gls{UE} to track the \gls{AoI} evolution, which can be formally expressed as:
\begin{align}\label{eq:aotevolution}
    \Delta_{n}(t+1) \coloneqq \min(\Delta_{n}(t), \Bar{\Delta})\cdot(1-\phi_{n}(t+1))+1.
\end{align}
Here, $\phi_{n}(t+1)$ is a joint Bernoulli random variable spanning the current and the next \glspl{TTI} $t$ and $t+1$ and models \gls{UE} selection, \gls{TB}-level instantaneous feedback and \gls{UE}'s packet replenishment (denoted by an indicator function with respect to the \gls{BSR} below). Formally, it is given by:
\begin{empheq}[left={\phi_n(t+1)}\coloneqq\empheqlbrace]{align}
   &1, ~\text{if } \beta_{n}(t) \land s_n(t) \land \mathbbm{1}_{\{b_{n}(t+1)>b_{n}(t)\}} = 1 \nonumber \\
   &0,\text{ otherwise.} \label{eq.jointbernoulli}
\end{empheq}
The significance of the joint condition in~\eqref{eq.jointbernoulli} is that the \gls{gNB} can deduce whether a \gls{UE}'s packet replenishment has occurred due to condition $(a)$ or $(b)$ from the above packet generation model. This also alleviates any need for explicit packet-level metadata other than the \gls{BSR} to be exchanged between \glspl{UE} and the \gls{gNB} for \gls{UL} \gls{UE} scheduling in \gls{XR} applications.

Besides \gls{AoI} modeling, we consider the time-averaged \gls{PAoI} as a metric to effectively evaluate the \acrshort{Qo(S)E} and the \gls{XR} \gls{KPI}. Denoted by $\Delta_{n, PAoI}$, the time-averaged \gls{PAoI} is the average of \gls{AoI} values right before $\phi_{n}(t) = 1$. Formally, $\Delta_{n, PAoI}$ can be expressed as:
\begin{align}\label{eq.paoimetric}
    \Delta_{n, PAoI} \coloneqq \frac{\sum_{\mathcal{T}} \Delta_{n}(t)}{\sum_{\mathcal{T}} \phi_{n}(t+1)}
\end{align}
Based on the above assumptions and modeling, we now present the optimization problem considered in this work.

%% file: problemformulation.tex
\section{Timely Throughput Maximization}
\label{sec.problemformulation}
Maximizing the number of satisfied \glspl{UE} in the network, termed \gls{XR} capacity, constitutes an important \gls{XR} \gls{KPI}, where a \gls{UE} is labeled satisfied if at least $99\%$ of its packet is successfully received within the \gls{PDB}~\cite{vpetrov2022xrover5g}. For the system model outlined in section~\ref{sec.networkmodel}, the above \gls{KPI} can be partially addressed by maximizing the system throughput through fair \gls{UE} scheduling. On top of this, \gls{UE} satisfaction can be incorporated using timely scheduling opportunities through the \gls{PAoI} metric defined in~\eqref{eq.paoimetric}. The throughput achieved in this manner is concisely referred to as the \textit{timely throughput} in this work and is formally expressed using the optimization problem below:
\begin{subequations}\label{eq.problem1}
    \begin{align}
      \textbf{P1:}&\maximize_{\pmb{\beta}(t); \forall{t}}~\limsup_{\substack{T\rightarrow\infty}}~\frac{1}{T} \sum_{\mathcal{T}}\sum_{\mathcal{N}} \beta_{n}(t) \cdot U_{\alpha}(Q_{n}(t)) \label{obj.p1} \\
      &~\text{subject to}~ \biggl\{\limsup_{\substack{T\rightarrow\infty}}~\Delta_{n, PAoI} \leq \Bar{D}\biggr\}_{n = 1}^{\Bar{N}}, \label{cons.p1c1}\\
      &~~~~~~~~\qquad\sum_{n \in \mathcal{N}}\lambda_n(t)\cdot\beta_n(t) \leq \Bar{\Lambda}, \label{cons.p1c2} \\
      &~~~~~~~~~~~~~~\pmb{\beta}(t) \in \{0,1\}^{\Bar{N}}, \forall n \in \mathcal{N}. \nonumber
    \end{align}
\end{subequations}
The problem posed in \eqref{eq.problem1} is an integer non-linear programming problem involving time-averaged expressions in~\eqref{obj.p1} and~\eqref{cons.p1c1}, along with instantaneous constraints in~\eqref{cons.p1c2}. The optimization variable $\pmb{\beta}(t) \in \{0,1\}^{\Bar{N}}$ is a binary vector indicating the co-scheduled \glspl{UE} at \gls{TTI} $t$. The function $U_{\alpha}(.)$ in~\eqref{obj.p1} denotes the well-known $\alpha$-fair utility function~\cite{muchida2009informationtheoretic} which is used to enforce fairness with respect to \gls{UE} scheduling. The interrelation between \glspl{UE}' \gls{PAoI}, $\Delta_{n, PAoI},~\forall n \in \mathcal{N}$ and the \gls{PDB}, $\Bar{D}$ is captured by the constraint set in~\eqref{cons.p1c1}. Such a coupling between \gls{PAoI} and \gls{PDB} is possible since the time instances when a \gls{UE}'s \gls{AoI} is reset to its default value signifies a packet replenishment. Furthermore, we consider multiple per-\gls{UE} constraints as opposed to a single network-wide utility function of \glspl{UE}' \gls{PAoI} to better reflect the \gls{XR} \gls{KPI}. Finally, the coupling constraint~\eqref{cons.p1c2} ensures that the total number of spatial layers served by \gls{gNB} does not exceed $\Bar{\Lambda}$. 

It can be observed that maximizing~\eqref{obj.p1} in isolation reduces \textbf{P1} to the classical fairness-enabled \gls{MU-MIMO} throughput maximization solely with respect to the \gls{PHY} layer aspects. However, inclusion of~\eqref{cons.p1c1} transforms \textbf{P1} into a cross-layer optimization problem delivering \textit{timely and meaningful} packet throughput to the \glspl{UE} within the \gls{PDB}. From a solution perspective, $\textbf{P1}$ seemingly looks like it can be decoupled across \glspl{UE} by relaxing the coupling constraint in~\eqref{cons.p1c2}. However, it is still not straightforward to decouple~\eqref{eq.problem1} since intra-cell interference acts as an implicit coupling constraint reflecting through $Q_n(t), \forall n \in \mathcal{N}$ computation. On top of this, the integer nature of the optimization variable $\pmb{\beta}(t)$ renders the above problem NP-hard. To effectively address these challenges, we propose a practically applicable, per-\gls{TTI} iterative heuristic algorithm to solve~\eqref{eq.problem1}, where the per-iteration decisions are evaluated using the \gls{PAoI}-weighted \gls{PF} metric.

%% file: proposedsolution.tex
\section{Proposed Scheduling Heuristic}
\label{sec.aoiwpf}

It is evident from section~\ref{sec.problemformulation} that a \gls{UE} scheduling scheme solving~\eqref{eq.problem1} should efficiently manage between \gls{UE} fairness, system throughput and \gls{PAoI} constraint satisfaction. The widely adopted \gls{PF}-metric-based schemes aim to strike a good balance between the first two attributes under different channel and interference conditions. As for~\eqref{eq.problem1}, utilizing the \gls{PF}-metric corresponds to the case when $\alpha = 1$ in $U_{\alpha}(.)$ and renders~\eqref{obj.p1} as maximizing a network-wide sum of per-\gls{UE} logarithmic utility of throughput at every \gls{TTI}, i.e., $\sum_{n \in \mathcal{N}} \log(Q_n(t))$. While fairness can be further improved by setting $\alpha > 1$, it can potentially impede timely scheduling opportunities to \glspl{UE} and impact the achievable \acrshort{Qo(S)E} and the \gls{XR} \gls{KPI}. This is because \glspl{UE} with worse channel conditions end up receiving a very high proportion of scheduling opportunities on an average in the \gls{TD} as $\alpha \rightarrow \infty$. Additionally, it can also result in intra-cell interference and eventually leads to packet replenishment at \glspl{UE} due to condition $(b)$ of the packet generation model from section~\ref{sec.networkmodel}. Considering all these, we employ a weighted \gls{PF}-metric-based scheduling heuristic algorithm to solve~\eqref{eq.problem1}, where the weights are derived as a function of \gls{PAoI} metric of each \gls{UE}.  

In order to maximize~\eqref{obj.p1} on a per-\gls{TTI} basis by setting $\alpha=1$ in $U_{\alpha}(.)$, i.e., $\sum_{n \in \mathcal{N}} \log(Q_n(t))$, one has to maximize $\sum_{n \in \mathcal{N}} \frac{Q_n(t)}{Q_{n, Avg}(t)}$~\cite{muchida2009informationtheoretic}. Here, $Q_n(t)$ is the per-\gls{UE} achievable throughput at \gls{TTI} $t$ from~\eqref{eq.uethroughput}, which is conditioned on the subset of co-scheduled \glspl{UE}, $\mathcal{N}^{\prime}(t)$. Furthermore, $Q_{n, Avg}(t)$ is a linear estimate of the per-\gls{UE} achieved throughput until \gls{TTI} $t$, for which the update rule is given by:
\begin{align}\label{eq.pfmetric}
    Q_{n, Avg}(t + 1) = (1-\tau)Q_{n, Avg}(t) + \tau\cdot \beta_n(t) \cdot Q_n(t).
\end{align}
Here, $\tau$ is a time constant, whose value is typically chosen to be in the order of $1000$ \glspl{TTI} (e.g. $\tau = \frac{1}{1000}$). We now extend this to~\eqref{eq.problem1} and accommodate the constraint set~\eqref{cons.p1c1} in the \gls{PF}-metric through a per-\gls{UE} multiplicative weight, $W_{n,PAoI}(t)$ scaling $Q_{n, Avg}(t)$ as defined below:
\begin{subequations}\label{eq.paoiweight}
    \begin{empheq}[left=W_{n, PAoI}(t)\coloneqq\empheqlbrace]{align}
    &(\Delta_{n, wa}(t))^{-\kappa},~~~~~\text{ if } \Delta_{n, wa}(t) \leq \Bar{D}  \label{eq.paoiweight1}\\
    &1-(\Delta_{n, wa}(t))^{-\kappa},~\text{otherwise.} \label{eq.paoiweight2}    
    \end{empheq}    
\end{subequations}    
Here, $\Delta_{n, wa}$ is the weighted average computed with respect to instantaneous \gls{AoI} $\Delta_n(t)$ and the running average of the \gls{PAoI} metric $\Delta_{n, PAoI}$ as:
\begin{align}\label{eq.weightedavg}
    \Delta_{n, wa}(t) := \theta \cdot \Delta_n(t) + (1-\theta) \cdot \sum_{t^{\prime} = 0}^{t}\Delta_{n, PAoI}(t^{\prime}).
\end{align}
The quantities $\kappa \in \mathbb{R}^{+}$ and $\theta \in [0,1]$ are interdependent tunable parameters. For instance, $\kappa$ assumes smaller values if $\Delta_{n, wa}(t)$ is prioritized and takes on higher values if $\sum_{t^{\prime} = 0}^{t}\Delta_{n, PAoI}(t^{\prime})$ is prioritized in~\eqref{eq.weightedavg}. Finally,~\eqref{eq.paoiweight} and~\eqref{eq.weightedavg} capture the R.H.S and L.H.S of~\eqref{cons.p1c1}, respectively.

Besides modeling the \gls{AoI}-weighted \gls{PF}-metric, we also incorporate~\eqref{cons.p1c2} in our proposed scheduling heuristic algorithm by leveraging the interconnection between $\Bar{\Lambda}$ and $\pmb{\Lambda}$. More specifically, we identify that~\eqref{cons.p1c2} essentially translates into an upper bound on the number of \glspl{UE} that can be co-scheduled in a \gls{TTI} $N(t) \leq \Bar{N}$, which can be approximated as follows: 
\begin{align}\label{eq.surrogatecond}
     N(t) := \ceil[\Big]{\frac{\Bar{\Lambda}}{\frac{1}{\Bar{N}}\sum_{n \in \mathcal{N}} \lambda_n(t)}},~\implies\pmb{1}^T\pmb{\beta}(t) \leq N(t), 
\end{align}

\begin{algorithm}
    \caption{Per-\gls{TTI} Timely Throughput Maximization}
    \label{alg.pseudocode}
    \begin{algorithmic}[1]
        \INIT $\tau, \kappa, \theta, \Bar{\Lambda}, \Bar{D}$.
        \INPUT \small{$\Delta_{n, wa}(t), \textbf{H}_n(f; t),\textbf{R}(i_f;t), \forall i_f \in \mathcal{I}_f,\forall f \in \mathcal{F},\forall n \in \mathcal{N}$}.
        \State Set ${\mathcal{N}}^{\prime}(t) \gets \varnothing$, ${\mathcal{K}}(t) \gets \varnothing$, and $Q_{sum}({\mathcal{N}^{\prime}}(t)) \gets 0$.
        \ForAll{$n \in \mathcal{N}$}
            \State ${\mathcal{K}}(t) \gets {\mathcal{N}^{\prime}}(t)$
            \If {$Q_{sum}({\mathcal{N}^{\prime}}(t) \cup \{n\}) > Q_{sum}(\mathcal{N}^{\prime}(t))$} 
                \State ${\mathcal{N}^{\prime}}(t) \gets {\mathcal{N}^{\prime}}(t) \cup \{n\}$
                \State $Q_{sum}(\mathcal{N}^{\prime}(t)) \gets Q_{sum}({\mathcal{N}^{\prime}}(t) \cup \{n\})$
            \EndIf
            \If {$|{\mathcal{N}^{\prime}}(t)| > N(t)~\text{OR}~\sum_{n\in\mathcal{N}^{\prime}(t)}\lambda_n(t) > \Bar{\Lambda}$}
            \State \Return $\mathcal{K}(t)$. \Comment Early stopping rule.
            \Else { Continue}
            \EndIf
        \EndFor
        \State \Return $\mathcal{N}^{\prime}(t)$. \Comment Equivalent to obtaining $\pmb{\beta}(t)$ in~\eqref{eq.problem1}.
    \end{algorithmic}
\end{algorithm}

The pseudocode for the proposed scheduling heuristic algorithm is presented in Algorithm~\ref{alg.pseudocode}, where the term $Q_{sum}({\mathcal{N}}^{\prime}(t)) \coloneqq \sum_{n \in \mathcal{N}^{\prime}(t)} \frac{Q_n(t)}{Q_{n, Avg}(t)\cdot W_{n, PAoI}(t)}$. The proposed scheduling heuristic algorithm serves the subset of \glspl{UE} which maximize $Q_{sum}$ at each \gls{TTI} $t$ i.e., it prioritizes the \glspl{UE} with a high $Q_n(t)$ value and a low value of the product $Q_{n, Avg} \cdot W_{n, PAoI}$. Thus, $W_{n, PAoI}$ in~\eqref{eq.paoiweight} guides the \gls{PF}-metric to prioritize a \gls{UE} until $\Delta_{n, wa}(t) \leq \Bar{D}$, beyond which it linearly approaches the classical \gls{PF}-metric. Thus, the proposed scheduling heuristic algorithm strikes a good balance between \gls{PAoI} constraint satisfaction, system throughput and \gls{UE} fairness.

%% file: numericalresults.tex
\section{Numerical Results}
\label{sec.numericalresults}
\begin{figure}
    \centering
    \includegraphics[width=0.7\textwidth]{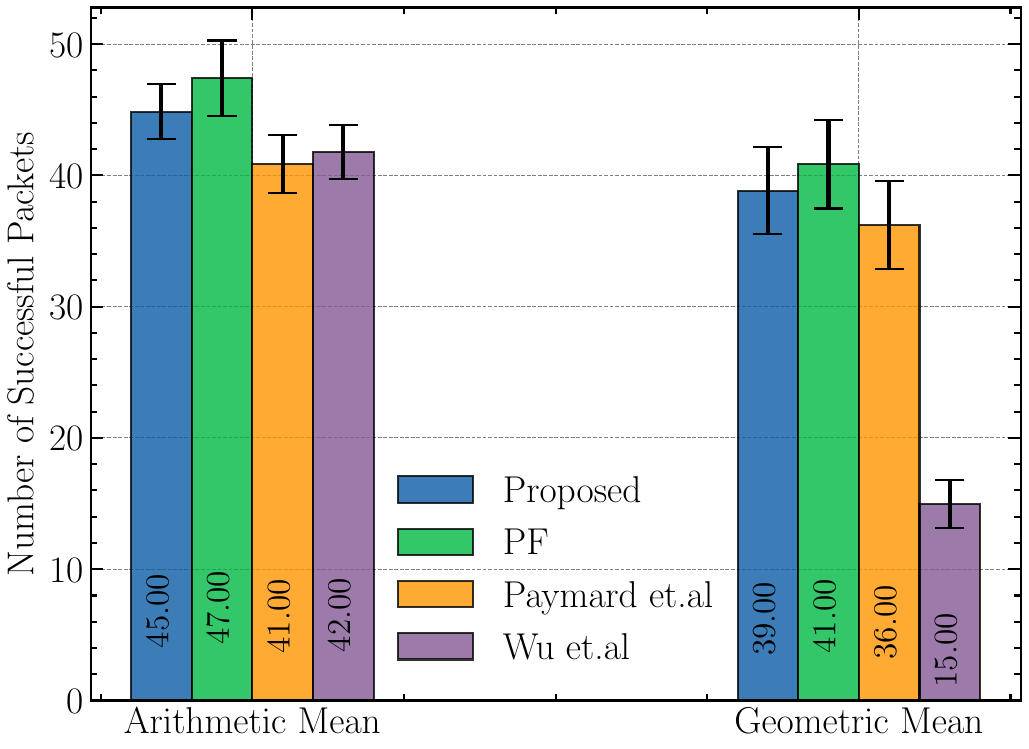}
    \caption{Goodput comparison of top $95\%$ \glspl{UE}}
    \label{fig.toptpt}
\end{figure}
\begin{figure}
    \centering
    \includegraphics[width=0.7\textwidth]{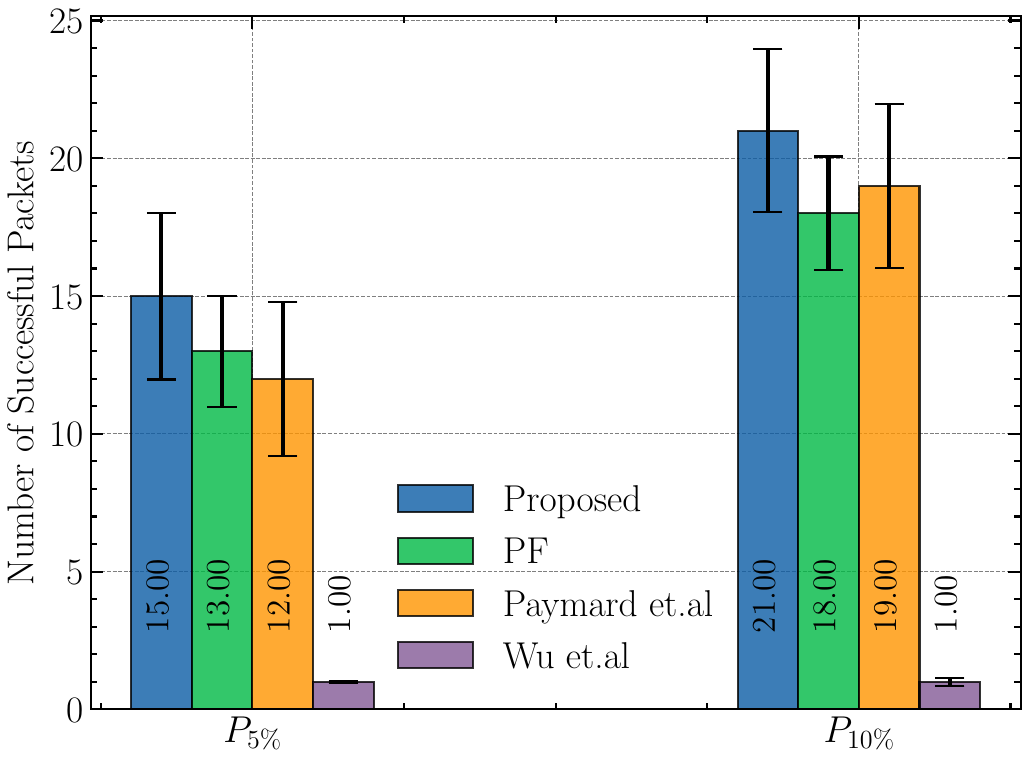}
    \caption{Goodput comparison of bottom $5\%$ and $10\%$ \glspl{UE}}
    \label{fig.bottomtpt}
\end{figure}
In this section, we evaluate the performance of Algorithm~\ref{alg.pseudocode} using a multi-cell (\gls{gNB} in this work), multi-link-level simulator written in NumPy (for channel estimation) and TensorFlow. We consider a single site with three \glspl{gNB}, where each \gls{gNB} is considered to serve $\Bar{N} = 10$ \glspl{UE} in \gls{MU-MIMO} mode. Each \gls{gNB} operates with $48~(3 \times 1~\text{subarray})$ antenna elements and is configured with $N_G = 16$-\gls{TRX} ($2\text{H} \times 4\text{V} \times 2\text{P}$). Furthermore, we consider the maximum spatial layers the \gls{gNB} can support at any \gls{TTI}, $\Bar{\Lambda} = 8$. The \glspl{UE} are configured with $N_U = 4$-\gls{TRX} ($2\text{H} \times 1\text{V} \times 2\text{P}$). For \gls{PHY} attributes, we consider the \gls{3GPP} $38.901$-Urban Micro (UMi) NLoS channel model with \gls{5G} pilots for channel estimation. Other relevant \gls{PHY} attributes such as \gls{UE} transmit power, \gls{RB}-settings, \gls{MCS} table and channel coding parameters etc., can be found in~\cite[sec. 5, tbl. I]{kpsrinath2024joint}). Furthermore, we also follow the baseline \gls{UL} power control, \gls{RB} allocation and \gls{UE} rank-selection from~\cite[sec. 5.A]{kpsrinath2024joint}). As for \gls{XR} application attributes, we consider the packet-size, $B_n(t) = \Bar{B} = 75000~\text{bits}~(9375~\text{Bytes}), \forall n \in \mathcal{N}$ for simplicity in evaluation. The \gls{PDB} value is selected as $\Bar{D}=30$ and consider an additional grace period of two \glspl{TTI} beyond \gls{PDB}. This is to provision \glspl{UE} to send their final remaining \glspl{TB} (if any) before packet replenishment. For the \gls{BSR} reporting, we follow~\cite[tbl. 6.1.3.1-1]{3gpp.38.321}. Furthermore, the \gls{AoI} clipping value is chosen as $\Bar{\Delta} = 100$. The tunable parameters from~\eqref{eq.paoiweight} and~\eqref{eq.weightedavg} are chosen as $\kappa = 2$ and $\theta = 0.5$. Finally, the simulations are run for $35$ individual drops and a total of $10^{6}$ \glspl{TTI}. As for the baselines, we consider the \gls{DRL} solution from~\cite{ccwu2021aoiaware5g} and adopt the classical \gls{PF}-metric and the weighted \gls{PF} metric in Algorithm~\ref{alg.pseudocode}. 

We begin our evaluation in Figs.~\ref{fig.toptpt} and~\ref{fig.bottomtpt} by comparing the goodput in terms of average number of packet replenishments at the \glspl{UE} with $90\%$ confidence interval. Here, the averaging is performed with respect to both \glspl{TTI} and the number of \glspl{UE} and is rounded off to the nearest integer. For the top $95\%$ \glspl{UE}, the proposed scheduling heuristic outperforms the baseline schemes in~\cite{ppouria2024pduset} and~\cite{ccwu2021aoiaware5g} with respect to both arithmetic and geometric means as shown in Fig.~\ref{fig.toptpt}. At the same time, the proposed scheduling heuristic achieves a comparable performance with respect to to the \gls{PF} baseline. On the other hand, for the bottom $5\%$ and $10\%$ or the cell-edge \glspl{UE}, the proposed heuristic consistently outperforms all the baselines as depicted in Fig.~\ref{fig.bottomtpt}. This shows that scaling the \gls{PF}-metric by \gls{PAoI} weights from~\eqref{eq.paoiweight} only slightly sacrifices the goodput of the top $95\%$ \glspl{UE} to accommodate the cell-edge \glspl{UE}. Furthermore, this is accomplished efficiently, since cell-edge \glspl{UE} transmit with low \gls{MCS} values (potentially causing interference) and require more scheduling opportunities on an average for successful packet deliveries.
\begin{figure}
    \centering
    \includegraphics[width=0.7\textwidth]{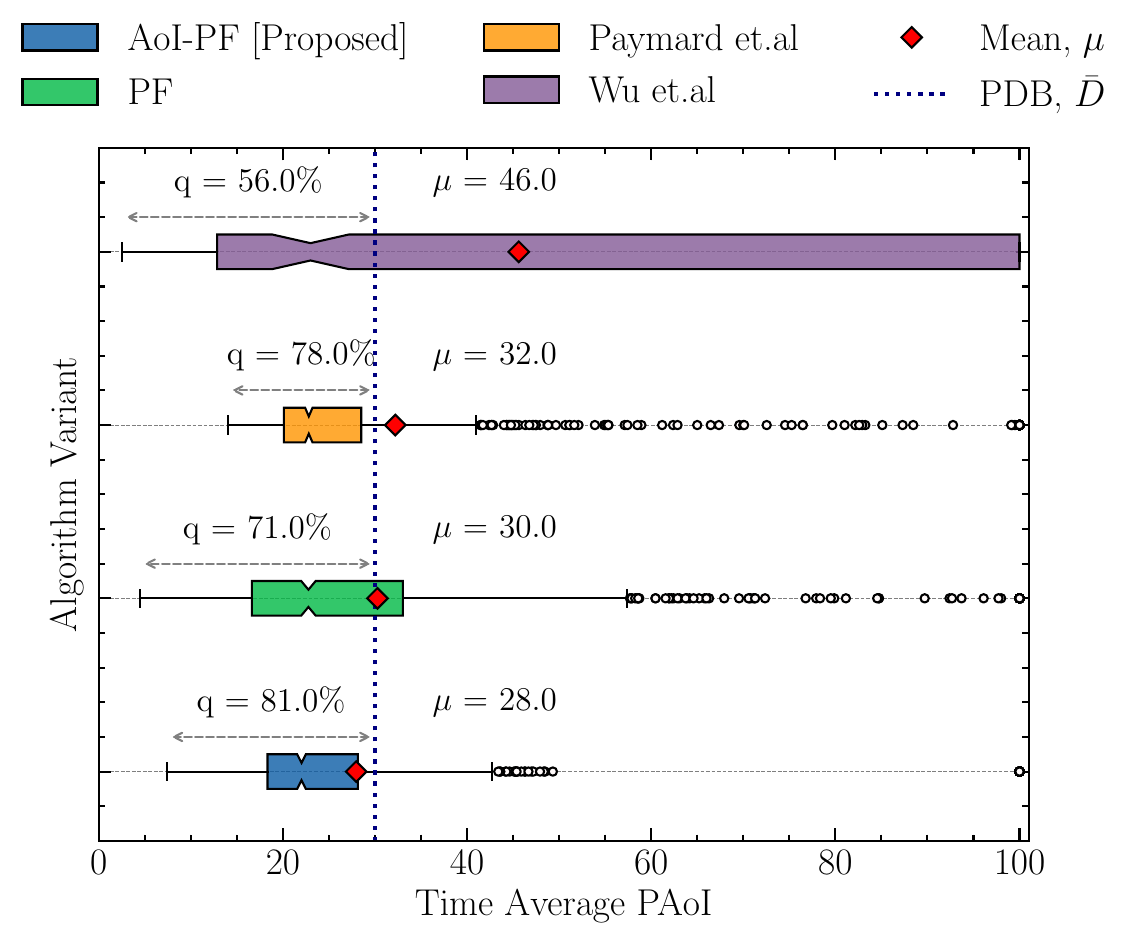}
    \caption{\gls{PAoI} distribution comparison}
    \label{fig.aoidist}
\end{figure}
\begin{figure}
    \centering
    \includegraphics[width=0.7\textwidth]{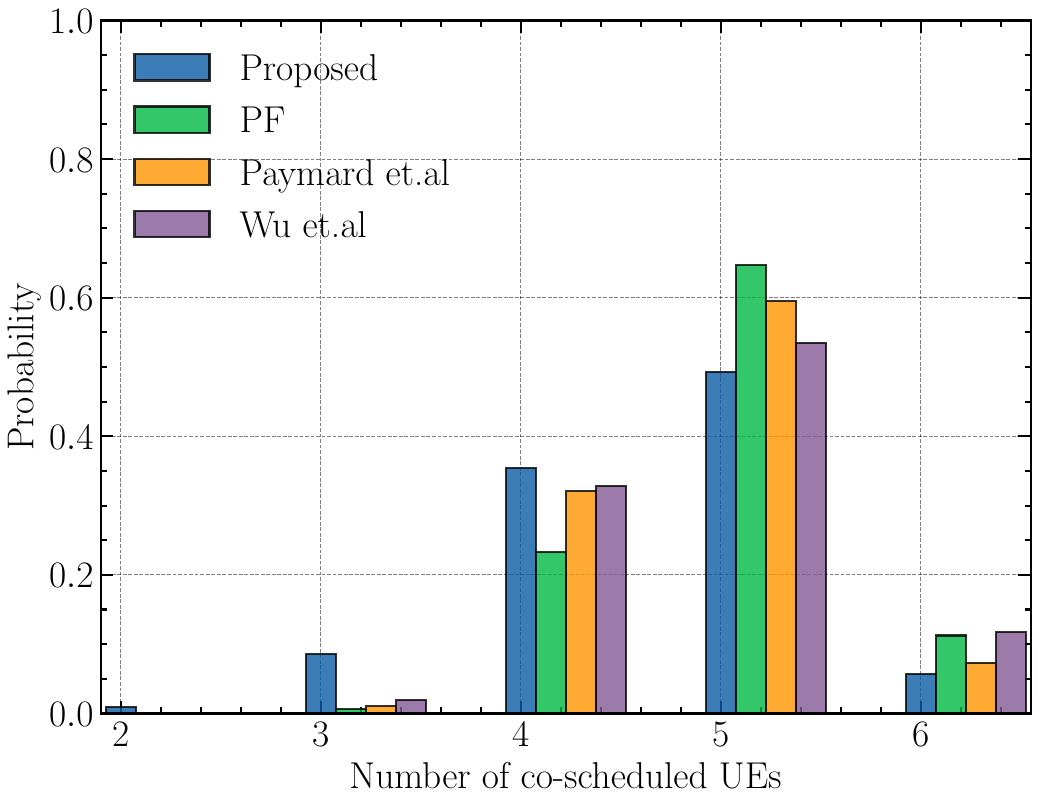}
    \caption{Empirical distribution of co-scheduled \glspl{UE}}
    \label{fig.coschedues}
\end{figure}

Next, we evaluate the \gls{XR} capacity achieved by the proposed scheduling heuristic and the baselines in Fig.~\ref{fig.aoidist}. More specifically, we compare the distributions of the time-averaged achieved \gls{PAoI} of the \glspl{UE} using a box plot figure. A box plot provides a compact distributional summary, where the data points are either partitioned into the quartiles or marked as outliers if they lie beyond the quartiles~\cite{boxplots}. The term $q$ in Fig.~\ref{fig.aoidist} represents the percentile of \glspl{UE} that satisfy constraint~\eqref{cons.p1c1}. Firstly, the proposed scheduling heuristic achieves the \textit{highest} \gls{XR} capacity of $q=81\%$. This is a significant improvement over the classical \gls{PF} scheduler, especially given that their goodput performance is comparable. Furthermore, a reasonable improvement in \gls{XR} capacity with respect to the baseline in~\cite{ppouria2024pduset} is still significant, since the proposed scheduling heuristic follows a signaling-free approach unlike~\cite{ppouria2024pduset}. Secondly, the proposed scheduling heuristic achieves the \textit{lowest} distributional mean of $\mu = 28$, which is well below the \gls{PDB}. This is because the outliers with respect to the proposed scheduling heuristic are closer to the fourth quartile compared to the baselines, whose outliers have longer tail distributions. Furthermore, despite classical \gls{PF} scheduler and the baseline in~\cite{ppouria2024pduset} achieving a mean close to the \gls{PDB}, their \gls{XR} capacity is still not maximized due to longer tail distributions. This asserts the need for per-\gls{UE} constraints in~\eqref{cons.p1c1} as opposed to the network-wide sum \gls{PAoI} minimization. Finally, it can be observed that the proposed scheduling heuristic outperforms the \gls{DRL} baseline in~\cite{ccwu2021aoiaware5g} with respect to both goodput and \gls{XR} capacity. This is because the \gls{DRL} baseline suffers from lack of explicit interference information due to \gls{MU-MIMO} and fails to adapt to \gls{XR} \glspl{KPI} in~\eqref{eq.problem1}. 

In Fig.~\ref{fig.coschedues}, we compare the \gls{pmf} of the number of co-scheduled \glspl{UE} with respect to each scheduling scheme. It can observed that the \gls{PAoI} weight from~\eqref{eq.paoiweight} regulates the number of \glspl{UE} scheduled when necessary. This is in contrast to the baselines, where the number of co-scheduled \glspl{UE} are concentrated more on the higher values.

%% file: conclusion.tex
\section{Conclusion}
\label{sec.conclusion}
In this work, we proposed an iterative \gls{UE} scheduling heuristic algorithm to solve the timely throughput maximization for \gls{UL} \gls{MU-MIMO} endowed with \gls{XR} \glspl{KPI}. Numerical simulations demonstrate that the proposed heuristic algorithm achieves the highest \gls{XR} capacity while not sacrificing the achieved throughput. We posit the possibility of utilizing \gls{DRL} techniques to reduce the computational complexity of the proposed approach.